\documentclass[prb,amsmath,amssymb,twocolumn,longbibliography]{revtex4-2}
\usepackage{graphicx}
\usepackage{dcolumn}
\usepackage{bm}
\usepackage{placeins}
\usepackage{rotating}
\usepackage{multirow}
\usepackage[dvipsnames,usenames]{xcolor}
\usepackage[english]{babel}
\usepackage{amssymb}
\usepackage{amsmath}
\usepackage[utf8]{inputenc}
\usepackage[normalem]{ulem}
\usepackage{float}

\newcommand{\be}{\begin{equation}}
\newcommand{\ee}{\end{equation}}
\newcommand{\bea}{\begin{eqnarray}}
\newcommand{\eea}{\end{eqnarray}}

\newcolumntype{M}[1]{>{\centering\arraybackslash}m{#1}}
\newcolumntype{N}{@{}m{0pt}@{}}

\def\a{\alpha}
\def\b{\beta}
\def\g{\gamma}

\def\d{\delta}
\def\D{\Delta}

\def\L{\Lambda}

\begin{document}
\widetext 

\title{Optimized effective potential forces with the plane-wave and pseudopotential method}
\author{Damian Contant}
\affiliation{Sorbonne Université, MNHN, UMR CNRS 7590, IMPMC, 4 place Jussieu, 75005 Paris, France}
\author{Maria Hellgren}
\affiliation{Sorbonne Université, MNHN, UMR CNRS 7590, IMPMC, 4 place Jussieu, 75005 Paris, France}
\date{\today}

\begin{abstract}
The optimized effective potential (OEP) approach has so far mainly been used in benchmark studies and for the evaluation of band gaps. In this work, we extend the application of the OEP by determining the analytical ionic forces within the plane-wave and pseudopotential framework. It is first shown that, due to the constrained optimization inherent to the OEP approach, an extra term needs to be added to the standard Hellmann-Feynman expression for the forces, whenever nonlocal pseudopotentials are employed. Computing this term for functionals based on Hartree-Fock and the hybrid PBE0 functional yields forces with excellent numerical accuracy. Furthermore, results for equilibrium geometries and vibrational frequencies on a set of molecules and solids confirm that the local exchange OEP is able to reproduce results obtained with the nonlocal exchange potential. Our work opens up the possibility to study lattice dynamics using advanced orbital functionals for describing exchange and correlation effects.
\end{abstract}
\maketitle
\section{\label{sec:Part1} Introduction}
In the field of computational physics, a wide variety of approaches for total energy and force calculations are available. Depending on the application, the accuracy-cost ratio can be an important factor to consider. In this regard, Kohn-Sham (KS) Density Functional Theory (DFT), with a well-chosen functional, is often the preferred approach \cite{HK64,KS65,parryang,Fiolhais2003}. 

Within KS-DFT, approximate functionals are sorted into different classes according to their level of description of the exchange-correlation (xc) energy \cite{Kohn1996,perdew2001,rappoport2008functional}. Starting from the local density approximation (LDA), improved functionals include a semilocal dependency through the gradient of the density, and are known as the generalized gradient approximations (GGAs). Adding a dependence on the KS kinetic energy density leads to the nonlocal meta-GGAs. Each of the above functional classes suffers, more or less, from self-interaction errors due to their approximate description of Hartree-Fock (HF) exchange \cite{Perdew1981}, a problem which can be mitigated with hybrid functionals that mix in a fraction of HF exchange into a semilocal functional \cite{becke1993,hse03}. In this way, the KS density matrix is introduced as an additional ingredient in the xc functional. An accurate description of correlation can be found with functionals based on the KS Green's function that depend on the full KS spectrum. Examples are functionals derived from many-body perturbation theory (MBPT) such as the random phase approximation (RPA) \cite{Ren2012}.
 
Meta-GGAs, hybrid and MBPT-based functionals all have an explicit dependency on the KS orbitals, rather than the density. As a consequence, the xc potential, i.e., the functional derivative of the xc energy with respect to the density, does not have an analytical expression. Self-consistent calculations are, therefore, often performed within the generalised KS framework \cite{Seidl1996,Garrick2020,Lebeda2023}, or by allowing for larger variational freedom. In the case of hybrid functionals this implies the use of an integral operator, the nonlocal exchange potential \cite{paier2006}, and, in the case of meta-GGAs, the use of a differential operator \cite{eichhellgren2014,perdewmetagga}. The RPA functional has a natural extension within MBPT. Free variations with respect to the many-body Green's function leads to a nonlocal and energy-dependent potential - the $GW$ self-energy \cite{klein1961,dahlen2006,hellgren2007}. Due to the high complexity of such calculations there are only a few reported in the literature \cite{holm1998,stan2006,caruso2012,kresse2018}.

In order to remain within the KS formulation, the variational freedom has to be restricted such that the orbitals are generated by a local multiplicative potential. Functional differentiation with respect to the density via the KS orbitals leads to an integral equation, known as the optimized effective potential (OEP) equation, for determining the xc potential numerically \cite{Sharp1953,Talman1976}. Tests on various systems have demonstrated that the OEP often gives total energies and densities in close agreement with results obtained within the generalised KS scheme \cite{Aashamar1979,KLI1992,Kim2017}. 
On the other hand, except for the highest occupied molecular orbital (HOMO) eigenvalue \cite{almbladh1985}, the spectrum will always be different. The KS virtual orbitals can, for example, be shown to provide a better description of the optical excitation energies \cite{Baerends2013}, but the fundamental gap is strongly underestimated, even with the exact KS potential \cite{Li1991,Perdew1985}.
One can show that the KS potential must jump with a constant when crossing integer particle numbers in the ensemble formulation of DFT \cite{Balduz1982}. Adding this constant, or the so-called derivative discontinuity correction, to the KS gap results in the true fundamental gap. Within hybrid functionals, the corrected KS gap is in good agreement with the gap obtained from the generalised KS scheme \cite{hellgren2021}, and same is expected to be true for meta-GGAs \cite{Lebeda2023}. Within RPA, it becomes equivalent to the gap within $G_0W_0$ theory \cite{Niquet2004,Gruning2006,Klimes2014}.

The numerical solution to the OEP equation has been the subject of numerous studies. A direct solution based on the inversion of the KS density response function is known to present numerical instabilities in Gaussian basis sets \cite{Gorling1999,gorlinghess2007,gorling2021}. These issues are almost absent from OEP calculations resorting to plane-waves \cite{stadele1999}, real space grids \cite{jiang2005,Makmal2009}, or spline basis sets \cite{hellgren2007,hellgren2008,hellgren2009,hellgren2010,hellgrengross2013}. Furthermore, direct minimization \cite{Yang2002} and iterative approaches \cite{Kummel2003_a,Kummel2003_b,Nguyen2014} that circumvent the inversion of the KS density response function have been developed, making routine OEP calculations feasible on a wide range of systems. All-electron studies have focused on the role of the core-valence interaction and concluded that the pseudopotential approximation is valid also for OEP calculations \cite{Engel2001,Sharma2005,Makmal2009,Makmal22009,betzinger2011}. A review on orbital-dependent functionals and their numerical aspects can be found in Ref. \onlinecite{Kummel2008}. 

Current applications of the OEP are mostly found in band structure calculations. Either for a direct comparison with experiment \cite{Trushin2018,trushin2019}, or as a starting point for $G_0W_0$ calculations \cite{Rinke_2005,qteish2006,Klimes2014,hellgren2021}. In principle, self-consistent OEP calculations also give access to the ionic forces, which are relevant for structural relaxation and phonon spectra. The prospect of using an advanced description of exchange and correlation for calculating these important properties has so far not been much explored. As of now, there are only a few works studying OEP forces on small molecules using Gaussian basis sets \cite{Wu2005,thierbach2020analytic_exx,thierbach2020analytic_rpa}. 

In the present study, we investigate the analytical OEP forces using the plane-wave basis set and norm-conserving pseudopotentials, and show that it is possible to achieve excellent numerical accuracy when applied to both molecules and solids.

The paper is organized as follows. In section II, we provide the mathematical details of the OEP approach. In section III, we discuss the OEP forces and their implementation using norm-conserving pseudopotentials. We also analyze the numerical accuracy achieved on various molecules and solids. Finally, in section IV, we exploit the calculated forces to determine equilibrium geometries, the vibrational modes of H$_2$O and $\a$-quartz, and the phonon dispersion of diamond. The conclusions are given in section V.

\section{\label{sec:Part2} Optimized effective potential}
Similarly to xc functionals depending explicitly on the density and its gradient, orbital-dependent functionals based on the OEP are designed to predict the density and the ground-state energy through a local effective KS potential. The self-consistent procedure does, however, present differences as the OEP KS potential does not have an analytical expression explicit in the orbitals. Instead, the numerical solution of the OEP integral equation is required as an intermediate step in each iteration towards self-consistency. 

The ground-state total energy within KS-DFT is written as
\begin{equation}
\displaystyle E_{\rm tot}= T_{\rm s} + E_{\rm Hxc}+ \int  v_{\rm ext}(\mathbf{r}) n(\mathbf{r}) \, d\mathbf{r},
\label{eq:DFT_Energy_Long}
\end{equation}
where $T_s$ is the kinetic energy of non-interacting electrons moving in an effective KS potential $v_{\rm eff}(\mathbf{r})$ such that
\begin{equation}
\displaystyle \biggl \{ -\frac{1}{2} \nabla^{2} + v_{\rm eff}(\mathbf{r})  \biggr \}  \varphi_{i}(\mathbf{r}) = \epsilon_{i} \varphi_{i}(\mathbf{r}).
\label{eq:DFT_KS_Equation}
\end{equation}
$E_{\rm Hxc}$ is the Hartree (H) and xc energy, and $v_{\rm ext}(\mathbf{r})$ is the external nuclear potential interacting with the electronic density $n(\mathbf{r})$. Each independent electron is described by a Kohn-Sham orbital $\varphi_i$ and has the energy $\epsilon_i$. The total energy is minimized when the effective potential is given by 
\begin{equation}
v_{\rm eff}(\mathbf{r})=v_{\rm ext}(\mathbf{r})  + v_{\rm Hxc}(\mathbf{r}) 
\label{eq:DFT_KS_potential}
\end{equation}
where 
\begin{equation}
v_{\rm Hxc}(\mathbf{r})=\frac{\d E_{\rm Hxc}}{\d n(\mathbf{r})}.
\label{eq:DFT_vhxc_potential}
\end{equation}

All terms in Eq. (\ref{eq:DFT_Energy_Long}) can be expressed explicitly in terms of the density except for the KS kinetic energy and the xc part of the Hxc energy. Since the functional dependence of the exact xc energy on the density is unknown, approximations are needed for its evaluation. The OEP method is relevant for xc functionals with an implicit dependence on the density via KS orbitals. In this case, the functional derivative in Eq. (\ref{eq:DFT_vhxc_potential}) is evaluated using the chain rule since the variation of the orbitals with respect to $v_{\rm eff}$ is easy to construct from linear response theory
\begin{equation}
\displaystyle \frac{\delta E_{\rm xc}}{\delta v_{\rm eff}(\mathbf{r})} = \int \frac{\delta n(\mathbf{r'})}{\delta v_{\rm eff}(\mathbf{r})} \, v_{\rm xc}(\mathbf{r'}) \, d\mathbf{r'}.
\label{eq:OEP_EXX_potential_Chi}
\end{equation}

Let us now focus on the exact-exchange (EXX) approximation. Within EXX, there is no correlation and the exchange energy functional is identical to the HF exchange energy \cite{KLI1992}

\begin{equation}
\displaystyle E_{\rm x}=- \frac{1}{4} \int \gamma(\mathbf{r},\mathbf{r}') v(\mathbf{r}-\mathbf{r'}) \gamma(\mathbf{r'},\mathbf{r}) \, d\mathbf{r}d\mathbf{r'}.
\label{EXXenergy}
\end{equation}
Assuming closed-shell systems, $\g$ is the first order spin-averaged reduced density matrix and $v$ is the Coulomb interaction. Free variations of the EXX total energy with respect to $\g$ yields the HF equation with the nonlocal HF exchange potential $V_{\rm x}(\mathbf{r},\mathbf{r'})=-\frac{1}{2}\,v(\mathbf{r}-\mathbf{r'}) \, \gamma(\mathbf{r'},\mathbf{r})$ \cite{parryang}. A variation with respect to a local effective potential corresponds to a minimization of the EXX total energy on a restricted domain of allowed orbitals, and the minimum is found when the exchange part of the effective potential obeys Eq.~(\ref{eq:OEP_EXX_potential_Chi}) \cite{Sharp1953,Talman1976}. We thus need to evaluate the functional derivative of $E_{\rm x}$ (Eq.~(\ref{EXXenergy})) with respect to $v_{\rm eff}$ 
\begin{equation}
\displaystyle \frac{\delta E_{\rm x}}{\delta v_{\rm eff}(\mathbf{r})} =\int \frac{\delta \gamma(\mathbf{r'},\mathbf{r''})}{\delta v_{\rm eff}(\mathbf{r})} V_{\rm x}(\mathbf{r'},\mathbf{r''})  \, d\mathbf{r'} d\mathbf{r''}.
\label{eq:OEP_EXX_potential_Lambda}
\end{equation}
Combining Eq.~(\ref{eq:OEP_EXX_potential_Chi}) and  Eq.~(\ref{eq:OEP_EXX_potential_Lambda}) yields the following integral equation for $v_{\rm x}(\mathbf{r})$
\begin{equation}
\begin{aligned}
\displaystyle \int \chi_{\rm s} (\mathbf{r}, \mathbf{r'}) v_{\rm x}(\mathbf{r'}) \, d\mathbf{r'}=\int \Lambda_s(\mathbf{r}, \mathbf{r'}, \mathbf{r''}) V_{\rm x}(\mathbf{r'},\mathbf{r''})  \, d\mathbf{r'} \, d\mathbf{r''},
\end{aligned}
\label{eq:OEP_Equation}
\end{equation}
known as the OEP equation. The variation of the electronic density with respect to $v_{\rm eff}$ is equal to the KS linear density response function and can be written explicitly as
\begin{equation}
\begin{aligned}
\displaystyle \chi_{\rm s}(\mathbf{r},\mathbf{r'}) & = 2\sum \limits_{i}^{\rm occ} \sum \limits_{j}^{\rm unocc} \frac{\varphi_{j}^{*}(\mathbf{r'}) \varphi_{j}(\mathbf{r}) \varphi_{i}^{*}(\mathbf{r}) \varphi_{i}(\mathbf{r'})}{\epsilon_{i}-\epsilon_{j}} \\
& + 2\sum \limits_{i}^{\rm occ} \sum \limits_{j}^{\rm unocc} \frac{\varphi_{i}^{*}(\mathbf{r'}) \varphi_{i}(\mathbf{r}) \varphi_{j}^{*}(\mathbf{r}) \varphi_{j}(\mathbf{r'})}{\epsilon_{i}-\epsilon_{j}}.
\end{aligned}
\label{eq:OEP_Chi_expression}
\end{equation}
The variation of $\gamma$ with respect to $v_{\rm eff}$ is given by 
\begin{equation}
\begin{aligned}
\displaystyle \Lambda_s(\mathbf{r},\mathbf{r'},\mathbf{r''}) & = 2\sum \limits_{i}^{\rm occ} \sum \limits_{j}^{\rm unocc} \frac{\varphi_{j}^{*}(\mathbf{r'}) \varphi_{j}(\mathbf{r}) \varphi_{i}^{*}(\mathbf{r}) \varphi_{i}(\mathbf{r''})}{\epsilon_{i}-\epsilon_{j}} \\
& +2 \sum \limits_{i}^{\rm occ} \sum \limits_{j}^{\rm unocc} \frac{\varphi_{i}^{*}(\mathbf{r'}) \varphi_{i}(\mathbf{r}) \varphi_{j}^{*}(\mathbf{r}) \varphi_{j}(\mathbf{r''})}{\epsilon_{i}-\epsilon_{j}}.
\end{aligned}
\label{eq:OEP_Lambda_expression}
\end{equation}
Both $\chi_s$ and $\L_s$ contain summations over occupied and unoccupied states. 

The OEP equation allows us to interpret the optimal local exchange potential as the potential that makes the perturbation $(V_{\rm x}-v_{\rm x})$, i.e. the perturbation that turns the KS equation into the HF equation, produce a vanishing first order density response \cite{Sham1983,casida2005}. 

The theory presented above can readily be generalized to other approximations based on nonlocal exchange such as hybrid functionals \cite{Kim2017,hellgren2021}. In the present work, we have used the PBE0 functional, which mixes in a fraction $\a=0.25$ of nonlocal exact exchange in the PBE functional. Contrary to the HF method, electronic correlation is present but remains described at the PBE level. 

A direct numerical solution of the OEP equation requires the construction of the response functions $\chi_s$ and $\L_s$, and the subsequent inversion of $\chi_s$. This procedure has been shown to work well on small systems but can give rise to numerical instabilities, in particular when Gaussian basis sets are used \cite{Gorling1999,gorlinghess2007,gorling2021}. With large basis sets, such as plane waves, it can instead be computationally demanding as it requires the summation over all unoccupied states and the manipulation of large matrices \cite{Nguyen2014}.

An iterative approach that avoids some of these problems was developed by Kümmel and Perdew \cite{Kummel2003_a,Kummel2003_b} and generalized to the plane-wave and pseudopotential framework by Nguyen et al \cite{Nguyen2014,Giannozzi2017}. The basic idea is to exploit the fact that the left and right hand sides of the OEP equation are linear density responses, $\delta n[v_{\rm x}]$ and $\delta n[V_{\rm x}]$, of the potentials $v_{\rm x}$ and $V_{\rm x}$, respectively. These responses can be calculated within Density Functional Perturbation Theory (DFPT), a framework developed for the calculation of phonon modes \cite{Baroni1987}. For a given trial potential $v_{\rm x}^i$, the two density responses will differ but their difference 

\begin{equation}
\Delta n^{i}(\mathbf{r})=\delta n[v^{i}_{\rm x}](\mathbf{r})-\delta n[V_{\rm x}](\mathbf{r})
\label{deltan}
\end{equation}
can be used to update $v_{\rm x}$ according to 
\begin{equation}
v_{\rm x}^{i+1}(\mathbf{r})=v_{\rm x}^{0}(\mathbf{r})+\sum_{m=1}^i\beta_m\Delta n^{m}(\mathbf{r}),
\label{eq:OEP_iterative_procedure}
\end{equation}
where $v_{\rm x}^{0}$ is the initial trial potential. In each iteration, the coefficients $\b_{m}$ are determined by minimizing the integral of $|\Delta n^{i+1}|$ and convergence is achieved when this value reaches a given threshold. The converged exchange potential is then used to update $v_{\rm eff}$ and the KS equation is solved non-self-consistently to generate a new set of orbitals that are used for solving the OEP equation another time. This cyclic procedure continues until the electronic density is found converged to a given threshold. An extension of this scheme to hybrid functionals and solids can be found in Refs. \onlinecite{hellgren2021,hellgren22021}. \\

\section{\label{sec:Part3} Analytical forces with OEP}
To understand the stability of a system, its set of nuclear forces needs to be computed. For a fixed configuration of the nuclei, each force $\mathbf{F}_{ I}$ describes the variation of the ground-state energy with respect to the position $\mathbf{R}_{ I}$ of a given nucleus $ I$ 
\begin{equation}
\displaystyle \mathbf{F}_{I} = - \frac{\partial E_{\rm tot}^\mathbf{R}}{\partial \mathbf{R}_{ I}}.
\label{eq:Forces_general}
\end{equation}
Following the notation in Ref.~\onlinecite{Baroni2001}, we denote the set of nuclear positions as $\mathbf{R}=\{\mathbf{R}_{I}\}$. At equilibrium geometry, the force exerted on each nucleus is zero. Calculating the second derivative of $E^\mathbf{R}_{\rm tot}$ with respect to $\mathbf{R}_{I}$ then yields interatomic force constants, which are necessary quantities for computing harmonic vibrational frequencies \cite{Pick1970}. 

The derivative in Eq.~(\ref{eq:Forces_general}) can be evaluated using the Hellmann-Feynman theorem (HFT) \cite{Hellman1937,Feynman1939}, which states that due to the stationary property of the total energy, it is sufficient to consider the explicit dependence on external parameters, here the nuclear positions. The HFT is valid in DFT and with the OEP method.

The total force becomes a sum of two components
\begin{equation}
\mathbf{F}_{ I}=\mathbf{F}_{I}^{\rm NN}+\mathbf{F}_{ I}^{\rm ext}.
\label{eq:totalforce}
\end{equation}
The first term is the derivative of the nuclear-nuclear potential energy 
\begin{equation}
\displaystyle \mathbf{F}_{ I}^{\rm NN} = - \frac{\partial E_{\rm NN}^{\mathbf{R}}}{\partial \mathbf{R}_{ I}} = - \frac{\partial}{\partial \mathbf{R}_{ I}}\sum \limits_{ J \neq I} \frac{Z_{ I} Z_{ J}}{|\mathbf{R}_{ I}-\mathbf{R}_{J}|},
\label{eq:Forces_internuclear}
\end{equation}
where $Z_I$ is the charge of nucleus $I$. The second term has its origin in the interaction between the nuclei and the electrons
\begin{equation}
\displaystyle \mathbf{F}_{ I}^{\rm ext}  = - \int n^{\mathbf{R}}(\mathbf{r}) \frac{\partial v_{\rm ext}^{\mathbf{R}}(\mathbf{r})}{\partial \mathbf{R}_{ I}} \, d\mathbf{r},
\label{eq:Forces_external}
\end{equation}
where
\begin{equation}
\displaystyle v_{\rm ext}^{\textbf{R}}(\textbf{r}) =  \sum \limits_{ I} \frac{Z_{I}}{|\mathbf{R}_{I}-\mathbf{r}|}.
\label{eq:External_potential}
\end{equation}
The calculation of $\mathbf{F}_{I}^{\rm ext}$ requires the knowledge of the electronic density $n^{\mathbf{R}}(\mathbf{r})$, obtained by performing a self-consistent calculation at fixed nuclear geometry $\mathbf{R}$. 

\subsection{Nonlocal pseudopotentials}
The force equations presented so far are valid in the context of all-electron calculations. However, in practice, pseudopotentials are often used to reduce the computational cost. In this approach, the core electrons are frozen and described by an effective interaction. In general, the pseudopotential is separated into two contributions \cite{Hamann1979,Martin2004}

\begin{equation}
\displaystyle v_{\rm ext}(\textbf{r},\textbf{r}') =  v_{\rm L}(\textbf{r})\delta(\textbf{r},\textbf{r}')+v_{\rm NL}(\textbf{r},\textbf{r}').
\label{eq:External_potential_total}
\end{equation}
The first term is fully local while the second is non-local both in the radial and the angular momentum dependence. This nonlocal contribution is written as a sum of projectors, which are functions of the spherical harmonics. The action of projectors is short-ranged as they are only defined within the core radius, i.e., the cutoff region \cite{Kleinman1982,Blochl1990,Vanderbilt1990}. Such a separable form for the pseudopotential allows an accurate reproduction of the scattering properties of the all-electron external potential. It also crucially improves the computational efficiency with respect to the size of the plane wave basis set used. 

A nonlocal external potential does not pose any problem for functionals depending explicitly on the density or the gradient of the density because the total energy is not only stationary with respect to the density, but also with respect to variations of the KS density matrix. The expression for the external force term, Eq.~(\ref{eq:Forces_external}), then only needs to be modified by replacing the density with the density matrix 
\begin{equation}
\displaystyle \mathbf{F}_{ I}^{\rm ext}  = - \int \gamma^{\mathbf{R}}(\mathbf{r},\mathbf{r'}) \frac{\partial v_{\rm ext}^{\mathbf{R}}(\mathbf{r},\mathbf{r}')}{\partial \mathbf{R}_{ I}} \, d\mathbf{r}d\mathbf{r}'.
\label{eq:Forces_external_nonlocal}
\end{equation}

However, the total energy with OEP functionals is not stationary with respect to the KS density matrix, but with respect to the effective local potential. This difference leads to an extra term in the expression for the OEP total force. To derive this term, let us look at the EXX total energy at fixed nuclear positions $\mathbf{R}$
\begin{equation}
\begin{aligned}
\displaystyle E^{\mathbf{R},{\rm EXX}}_{\rm tot} & = 2\sum_{i}^{\rm occ} \epsilon^{\mathbf{R}}_{i} - \int v_{\rm Hx}^{\mathbf{R}}(\mathbf{r}) n^{\mathbf{R}}(\mathbf{r}) \, d\mathbf{r}  \\
& + \frac{1}{2} \int n^{\mathbf{R}}(\mathbf{r}) v(\mathbf{r}-\mathbf{r'}) n^{\mathbf{R}}(\mathbf{r'}) \, d\mathbf{r} d\mathbf{r'} \\
& - \frac{1}{4} \int \gamma^{\mathbf{R}}(\mathbf{r},\mathbf{r}') v(\mathbf{r}-\mathbf{r'}) \gamma^{\mathbf{R}}(\mathbf{r'},\mathbf{r}) \, d\mathbf{r} d\mathbf{r'}\\
& + E_{\rm NN}^{\mathbf{R}},
\end{aligned}
\label{eq:Energy_HF}
\end{equation}

and take the derivative with respect to $\mathbf{R}_I$. We get straightforwardly
\begin{equation}
\begin{aligned}
\displaystyle \mathbf{F}^{\rm EXX}_{ I} &=  \mathbf{F}^{\rm NN}_{ I}+ \mathbf{F}^{\rm ext}_{ I} +\mathbf{\D F}^{\rm EXX}_{I},
\end{aligned}
\label{eq:Forces_total_EXX}
\end{equation}
where $\mathbf{F}^{\rm ext}_{ I} $ is given by Eq.~(\ref{eq:Forces_external_nonlocal}) and 
\begin{equation}
\begin{aligned}
\displaystyle \mathbf{\D F}^{\rm EXX}_{I} &=  -\int v_{\rm x}^{\mathbf{R}}(\mathbf{r}) \frac{\partial n^{\mathbf{R}}(\mathbf{r})}{\partial \mathbf{R}_I} \, d\mathbf{r} \\
 &  +\int  V^{\mathbf{R}}_{\rm x}(\mathbf{r},\mathbf{r'})\frac{\partial \gamma^{\mathbf{R}}(\mathbf{r},\mathbf{r}')}{\partial \mathbf{R}_I} \, d\mathbf{r} d\mathbf{r'}.
\end{aligned}
\label{eq:Forces_delta_EXX}
\end{equation}
If the external potential was fully local, $\mathbf{\D F}^{\rm EXX}_{I} $ would vanish at self-consistency thanks to the OEP equation, Eq.~(\ref{eq:OEP_Equation}), being fulfilled. However, when the derivative of $n^{\mathbf{R}}$ and $\gamma^{\mathbf{R}}$ is taken via a nonlocal potential, the two terms in Eq.~(\ref{eq:Forces_delta_EXX}) are not guaranteed to cancel. To see this, let us look more closely at the derivative of the density with respect to the nuclear positions
\begin{equation}
\begin{aligned}
\displaystyle  \frac{\delta n^{\mathbf{R}}(\mathbf{r})}{\delta v_{\rm eff}^{\mathbf{R}}(\mathbf{r'},\mathbf{r''})} \frac{\partial v_{\rm eff}^{\mathbf{R}}(\mathbf{r'},\mathbf{r''})}{\partial \mathbf{R}_I}  =\frac{\delta n^{\mathbf{R}}(\mathbf{r})}{\delta v_{\rm eff}^{\mathbf{R}}(\mathbf{r'},\mathbf{r''})}\times \\
&\!\!\!\!\!\!\!\!\!\!\!\!\!\!\!\!\!\!\!\!\!\!\!\!\!\!\!\!\!\!\!\!\!\!\!\!\!\!\!\!\!\!\!\!\!\!\!\!\!\!\!\!\!\!\!\!\!\!\!\!\!\!\!\!\!\!\!\!\!\!\!\!\!\!\!\!\!\!\!\!\!\!\!\!\!\!\!\!\!\!\!\!\!\!\!\!\!\!\!\!\!\!\left[\left(\frac{\partial v_{\rm L}^{\mathbf{R}}(\mathbf{r'})}{\partial \mathbf{R}_I}+\frac{\partial v_{\rm Hxc}^{\mathbf{R}}(\mathbf{r'})}{\partial \mathbf{R}_I}\right)\d(\mathbf{r'},\mathbf{r''})+\frac{\partial v_{\rm NL}^{\mathbf{R}}(\mathbf{r'},\mathbf{r''})}{\partial \mathbf{R}_I}\right].
\end{aligned}
\label{eq:Forces_delta_EXX2}
\end{equation}
The derivatives via the local potentials ($v_{\rm L}^{\mathbf{R}}$ and $v_{\rm Hxc}^{\mathbf{R}}$) involve the standard non-interacting KS density response function (see Eq.~(\ref{eq:OEP_Chi_expression})). Therefore, when the OEP equation is fulfilled, they cancel exactly the corresponding contributions coming from the derivative of the density matrix with respect to the nuclear positions in Eq.~(\ref{eq:Forces_delta_EXX}). However, the derivative via the nonlocal potential ($v_{\rm NL}^{\mathbf{R}}$) requires the three-argument non-interacting KS density response function. The OEP equation can thus not be used and we are left with the following extra force term to evaluate
\begin{equation}
\begin{aligned}
\displaystyle \mathbf{\D F}^{\rm EXX}_{I} &=  -\int v_{\rm x}^{\mathbf{R}}(\mathbf{r})\frac{\delta n^{\mathbf{R}}(\mathbf{r})}{\delta v_{\rm eff}^{\mathbf{R}}(\mathbf{r'},\mathbf{r''})} \frac{\partial v_{\rm NL}^{\mathbf{R}}(\mathbf{r'},\mathbf{r''})}{\partial \mathbf{R}_I}  d\mathbf{r}d\mathbf{r'}d\mathbf{r''} \\
 & \!\!\!\!\!\!\!\!\!\!\!\!\!\!\!\!\!\!\!+ \int V^{\mathbf{R}}_{\rm x}(\mathbf{r},\mathbf{r'})\frac{\delta \gamma^{\mathbf{R}}(\mathbf{r},\mathbf{r}')}{\delta v_{\rm eff}^{\mathbf{R}}(\mathbf{r''},\mathbf{r'''})}\frac{\partial v_{\rm NL}^{\mathbf{R}}(\mathbf{r''},\mathbf{r'''})}{\partial \mathbf{R}_I} \, d\mathbf{r} d\mathbf{r'}d\mathbf{r''}d\mathbf{r'''}.
\end{aligned}
\label{eq:Forces_delta_EXX3}
\end{equation} 
We note that, since the nonlocal part of the 
external potential is fixed in the self-consistent procedure, i.e., it does not depend on the orbitals, only the bare responses are needed. The computational cost of this extra force term, calculated for the complete set of ions, is, therefore, estimated to be similar to a single iteration of the self-consistent procedure.

The extra force term due to the nonlocal pseudopotential will appear for any functional based on the OEP approach. In this work we have focused on hybrid functionals and norm-conserving pseudopotentials, and implemented Eq.~(\ref{eq:Forces_delta_EXX3}) at the end of the self-consistent cycle. This was done within a modified version of the OEP implementation in the \texttt{ACFDT} (Adiabatic Connection Fluctuation Dissipation Theorem) package of the \texttt{QUANTUM ESPRESSO} distribution \cite{Nguyen2009,Nguyen2014,Giannozzi2017,hellgren2021}. In order to evaluate the change in the KS orbitals with respect to variations in the ionic positions, we have imported and adapted routines based on DFPT 
from the \texttt{PHonon} package. The internuclear and external force terms (Eqs.~(\ref{eq:Forces_internuclear}) and (\ref{eq:Forces_external_nonlocal})) are general and both already implemented in the \texttt{PWscf} package. 

With the knowledge of analytical forces, the design of a structural optimization tool for the OEP method can be devised. We have completed it by adapting the existing Broyden-Fletcher-Goldfarb-Shanno algorithm \cite{Broyden1970,Fletcher1970,Goldfarb1970,Shanno1970} implemented within \texttt{PWscf}.
\subsection{Numerical test}
As discussed in section II, the OEP self-consistent procedure runs two intertwined parts. One solves the Kohn-Sham equation non-self-consistently to retrieve the electronic density and the corresponding KS orbitals of the ground-state, while the other solves the OEP equation iteratively to generate the local EXX potential. The whole procedure is initialized with a good starting guess for the KS potential. We have found the PBE approximation to be a convenient choice in this regard. In addition to the plane wave basis set cutoff, there are two parameters that control the accuracy of the final results. The accuracy of the iterative solution to the OEP equation is determined by setting a threshold for $\int |\Delta n|$ (see Eq.~(\ref{deltan})), and the accuracy of the self-consistent OEP potential is determined by setting a threshold for the difference in KS densities in-between successive cycles.

\begin{figure}[H]
\includegraphics[scale=0.37,angle=0]{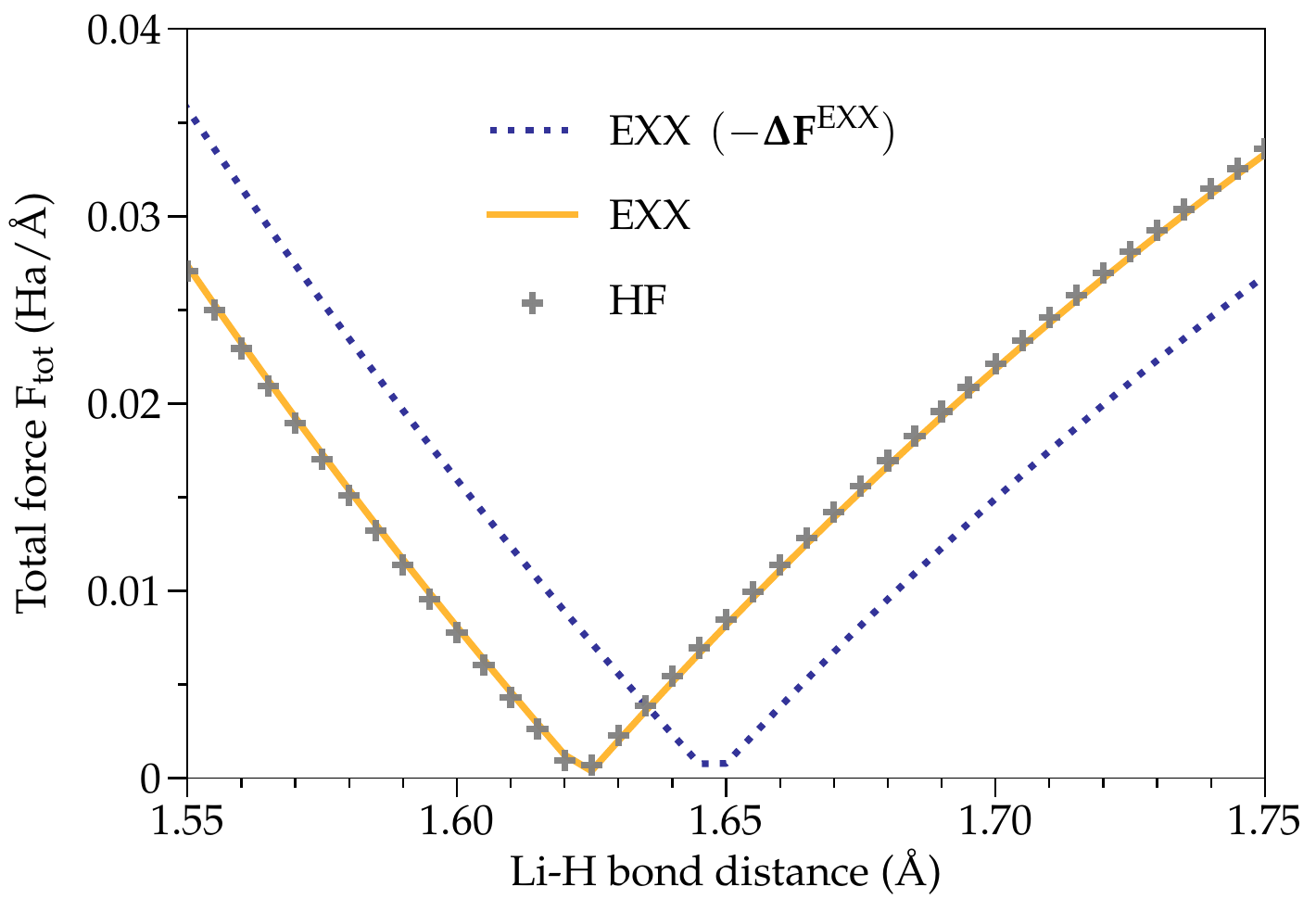}
\caption{\label{fig:Figure_LiH_HF_EXX} Change in the analytical total force $\mathrm{F}_{\rm tot}$ (defined in Eq.~(\ref{eq:Total_force_norm})) of the LiH molecule upon variations of its intramolecular bond distance. The HF method is compared to EXX with and without the extra force term in Eq.~(\ref{eq:Forces_delta_EXX3}).}
\end{figure}

In principle, one should use pseudopotentials optimized for the specific functional used \cite{Engel2001}. However, since our objectives are to study the numerical precision and to compare OEP to generalized KS calculations we have settled for PBE optimized norm-conserving Vanderbilt (ONCV) pseudopotentials \cite{Hamann2013}. Independently of the functional used to optimize the pseudopotential they will, in general, contain nonlocal projectors and the extra OEP force term in Eq.~(\ref{eq:Forces_delta_EXX3}) is necessary to include.

We first tested our implementation of the OEP forces on a simple system, the LiH molecule. We used a simulation cell of 20 Bohr and a plane-wave basis set with a kinetic energy cutoff of 80 Ry.
The quantity we are interested in is the so-called "total force", which is defined as 
\begin{equation}
\mathrm{F}_{\rm tot} = \sqrt{\sum_{I=1}^{M} \sum_{\alpha=1}^{3} (F_{I}^{\alpha})^{2}} 
\label{eq:Total_force_norm}
\end{equation}
where the $\a$-index runs over the Cartesian components of the force on each of the $M$ ions. 
The quantity $\rm F_{\rm tot}$ is a positive-definite scalar able to describe the global behaviour of the system. Its value decreases upon approaching an energy minimum, and is zero at equilibrium.

\begin{figure}[H]
\begin{center}
\includegraphics[scale=0.37,angle=0]{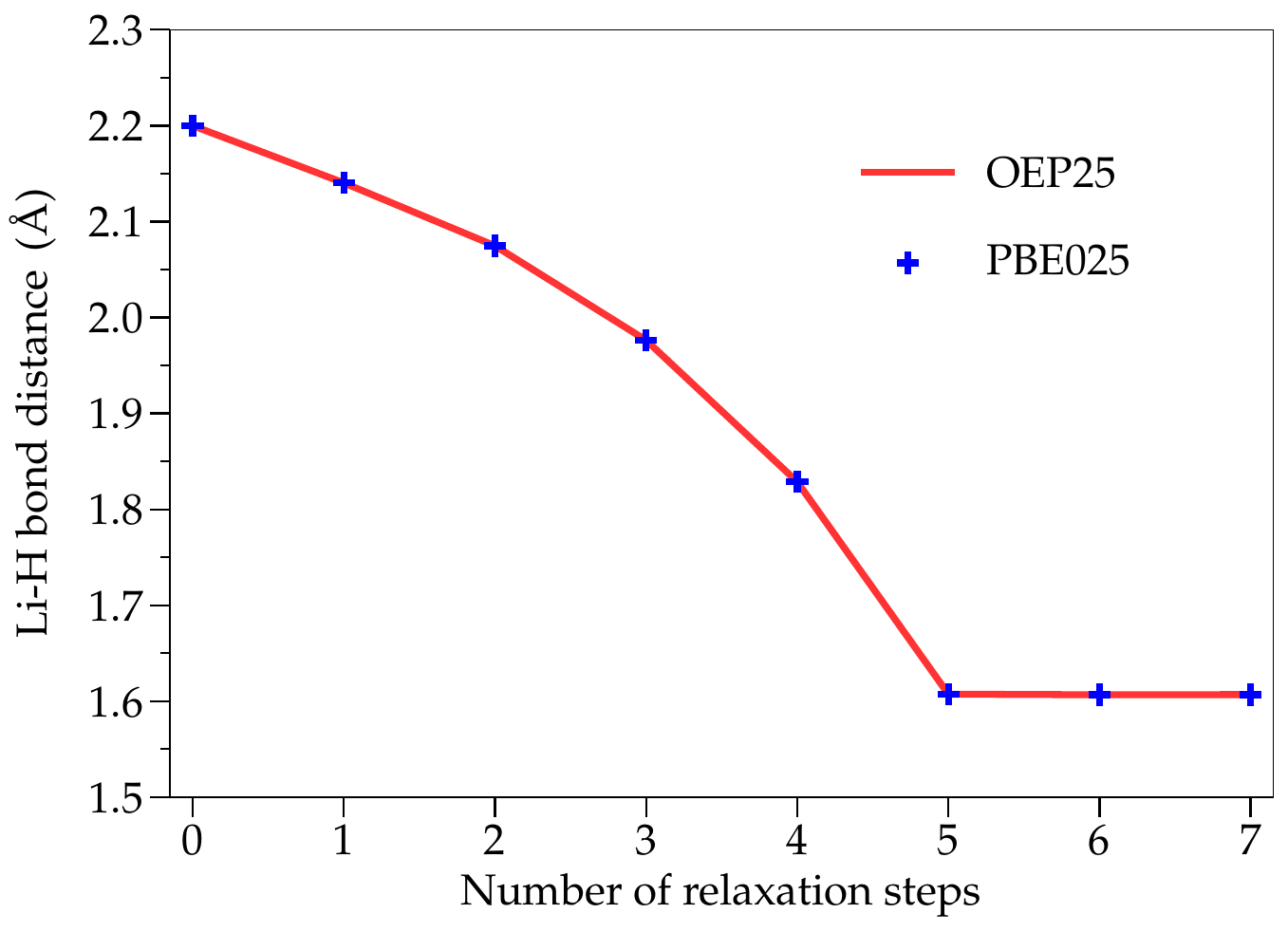}
\end{center}
\caption{\label{fig:Figure_LiH_PBE025_OEP25} Evolution of the intramolecular bond distance of the LiH molecule after each iteration of the geometry relaxation procedure. The nonlocal PBE025 functional is compared to the local OEP25.}
\end{figure}

In Fig. \ref{fig:Figure_LiH_HF_EXX}, the total force is plotted as a function of the Li-H intramolecular bond distance within the HF and EXX approximations. The EXX calculations have been performed with and without the inclusion of the extra OEP force term, $\D \mathbf{F}^{\rm EXX}$. Without this term, EXX predicts an equilibrium bond distance at 1.645 \AA \, as compared to 1.625 \AA \, for HF. The difference of 0.020 \AA \, corresponds to a difference in $\rm F_{\rm tot}$ of 0.008 Ha/\AA. On the other hand, if we looked at the total EXX energy as a function of bond distance, we would find the same geometry at equilibrium as with HF. This inconsistency can be explained by the missing OEP force term, previously identified. Including it in the calculation of $\rm F_{\rm tot}$ gives a good agreement between the EXX and HF total forces on the LiH molecule. Given that the importance of including nonlocal projectors in pseudopotentials increases with the number of electrons, we expect that $\D \mathbf{F}^{\rm EXX}$ will become even more important for heavier elements.

Having verified that the OEP forces come out accurately, we then optimized the geometry of the LiH molecule. Since the HF approximation is not expected to give a good equilibrium geometry, we used PBE0 with 25\% of exact exchange (PBE025). We compared the fully nonlocal PBE025, already implemented, to the corresponding local OEP version, which we call OEP25. In Fig. \ref{fig:Figure_LiH_PBE025_OEP25}, we see that starting from a Li-H bond length of 2.200 \AA, the convergence of the geometry with OEP25 follows the same pattern as PBE025. The relaxation steps are similar, with both methods returning, after only five iterations, the same intramolecular bond distance of 1.607 \AA. 

We will now present a more comprehensive study of the accuracy of OEP forces by comparing the analytical force, F$_{\rm A}$, as obtained from Eq.~(\ref{eq:Forces_total_EXX}), with the numerical force, F$_{\rm N}$, calculated by finite difference, using the five-point stencil formula with a step size of 0.01 \AA. Given that energies converge faster than forces thanks to error cancellation, numerical forces also converge faster than analytical ones. Indeed, F$_{\rm A}$ is calculated by omitting certain contributions that are zero only at perfect self-consistency. The absolute difference $|{\rm F}_{\rm A} - {\rm F}_{\rm N}|$ can therefore be viewed as an estimation of the error on the analytical forces calculated and the quality of the self-consistent procedure.
\begin{figure}[t]
\includegraphics[scale=0.48,angle=0]{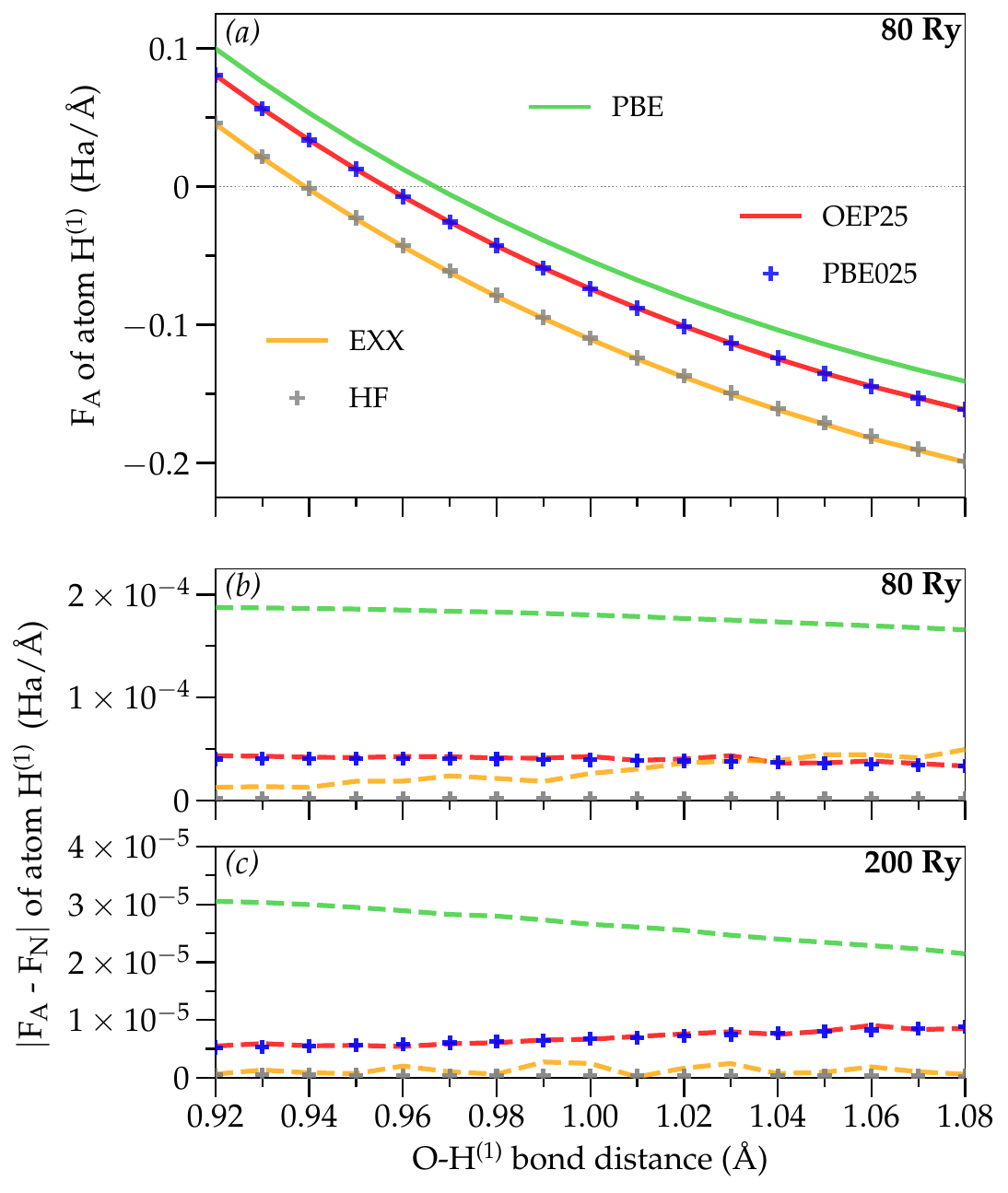}
\caption{\label{fig:Figure_H2O_methods} (a) Change in the analytical force F$_{\rm A}$ exerted on one of the hydrogen atoms of the H$_2$O molecule along the bond it forms with the O atom, according to different approximations. The corresponding numerical force F$_{\rm N}$ is not plotted as it is indistinguishable from F$_{\rm A}$ on the scale of the figure. (b) Absolute difference between F$_{\rm A}$ and F$_{\rm N}$ forces at fixed cutoff of 80 Ry and (c) 200 Ry.}
\end{figure}
\begin{figure}[t]
\includegraphics[scale=0.48,angle=0]{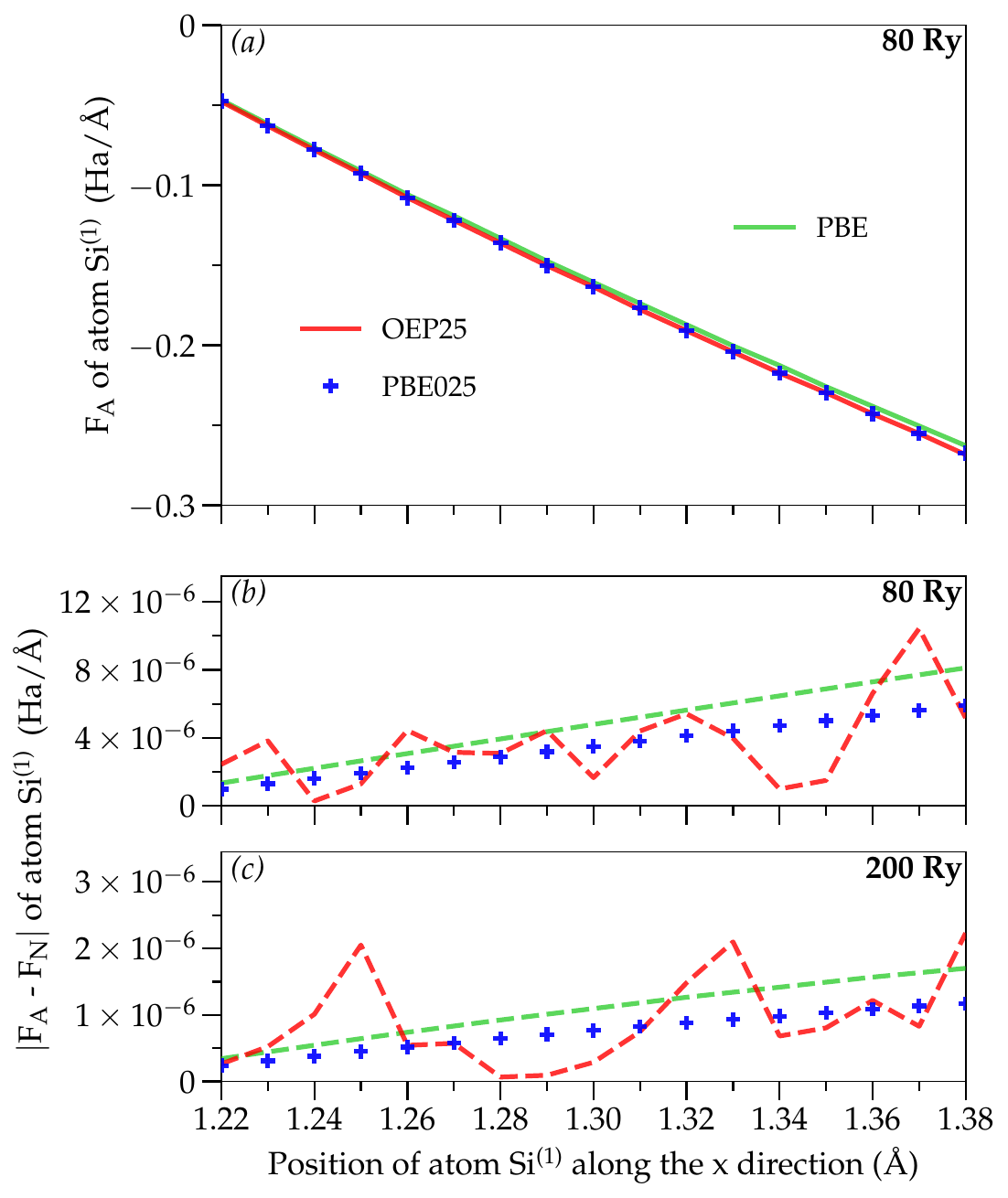}
\caption{\label{fig:Figure_SiO2_methods} (a) Change in the analytical force F$_{\rm A}$ exerted on one of the silicon ion of the $\alpha$-SiO$_2$ quartz phase upon variations of its position along the Cartesian x-axis, according to different approximations and at a plane-wave cutoff of 80 Ry. The corresponding numerical force F$_{\rm N}$ is not plotted as it is indistinguishable from F$_{\rm A}$ on the scale of the figure. (b) Absolute difference between F$_{\rm A}$ and F$_{\rm N}$ forces at fixed cutoff of 80 Ry and (c) 200 Ry.}
\end{figure}
We started by studying the water molecule H$_2$O (see Fig. \ref{fig:Figure_H2O_methods}) using five different approximations: PBE, HF, EXX, PBE025, and OEP25. All data have been obtained using a simulation cell of 25 Bohr, and results for 80 Ry and 200 Ry plane-wave cutoff are compared. The threshold on the energy convergence of PBE, HF, and PBE025 was set to lowest possible value to ensure high accuracy of the numerical forces. The same was done for the two different thresholds used by EXX and OEP25 (see Sec. II).
Regarding the atomic structure, the angle between the two O-H bonds has been fixed to 104.3°. Only the O-H$^{(1)}$ bond distance is varied while the length of O-H$^{(2)}$ is set to 0.97 \AA. These parameters define a geometry for the water molecule that is close to the PBE equilibrium geometry. By performing several test calculations at different geometries, we found that the choice of starting geometry does not impact the results we obtained on H$_2$O. 

In Fig. \ref{fig:Figure_H2O_methods}$(a)$, we plot the analytical force on atom H$^{(1)}$ along the O-H$^{(1)}$ bond distance as the bond is stretched. A smooth behaviour is observed for every method. Similarly to LiH, the effect of using a local OEP potential as an approximation to the nonlocal Fock exchange potential appears very small for both HF and PBE025.
\begin{table*}[ht!]
\caption{\label{tab:Tab_H2O} Structure parameters and vibration frequencies of the H$_2$O molecule calculated at equilibrium geometry with various methods. The O-H bond distance and the H-O-H bond angle are given. The frequencies have been calculated for the infrared active modes in-plane scissoring $\delta_{\text{H-O-H}}$, symmetric stretching $\nu^{s}_{\text{O-H}}$, and asymmetric stretching $\nu^{as}_{\text{O-H}}$. Results obtained in the literature with CCSD(T)/TZ(2df,2pd) \cite{Thomas1993} as well as experimental results \cite{Clabo1988} are also presented for reference.}
\vspace{5pt}
\centering
\begin{tabular}{M{3.5cm} M{2.5cm} M{2.5cm} M{2.5cm} M{2.5cm} M{2.5cm} N}
    \toprule
    \textbf{} & \textbf{d(O-H) (\AA)} & \textbf{$\Theta$(H-O-H) (°)} & \textbf{$\delta_{\textbf{H-O-H}}$ (cm$^{-1}$)} & \textbf{$\nu^{s}_{\textbf{O-H}}$ (cm$^{-1}$)} & \textbf{$\nu^{as}_{\textbf{O-H}}$ (cm$^{-1}$)} &\\
    \hline
    \textbf{PBE} & 0.9668 & 104.38 & 1597 & 3703 & 3816 &\\
    \textbf{PBE025} & 0.9558 & 104.91 & 1640 & 3857 & 3971 &\\
    \textbf{OEP25} & 0.9558 & 104.91 & 1639 & 3858 & 3972 &\\
    \textbf{HF} & 0.9385 & 106.06 & 1760 & 4113 & 4221 &\\
    \textbf{EXX} & 0.9380 & 106.17 & 1757 & 4123 & 4233 &\\
    \textbf{CCSD(T)} \cite{Thomas1993} & 0.9594 & 104.2 & 1650 & 3835 & 3944 &\\
    \textbf{Harmonic expt} \cite{Clabo1988} & 0.9572 & 104.52 & 1649 & 3832 & 3943 &\\
    \textbf{Anharmonic expt} \cite{Clabo1988} & 0.9572 & 104.52 & 1595 & 3657 & 3756 &\\ \toprule 
\end{tabular}
\end{table*}
In Fig.~\ref{fig:Figure_H2O_methods}$(b)$, the difference between analytical and numerical forces is calculated at a cutoff of 80 Ry. For every method, the scale of the error on F$_{\rm A}$ is found to be relatively small. The performance of the hybrid PBE025 functional is one order of magnitude better than PBE, and three orders of magnitude better with HF. If we now analyze the performance of the OEP methods, we notice that OEP25 produces a similar error to PBE025, which on average is about $4 \times 10^{-5}$ Ha/\AA \, at 80 Ry. On the other hand, EXX fails to deliver the exceptionally small error seen with HF. While HF returns force differences of the order of $10^{-7}$ Ha/Å, EXX results are closer in magnitude to PBE025 and OEP25. The difference observed between EXX and HF is probably related to the threshold used to solve the OEP equation. In Fig.~\ref{fig:Figure_H2O_methods}$(c)$ we present the same curves as in Fig. \ref{fig:Figure_H2O_methods}$(b)$ but with a higher plane-wave cutoff of 200 Ry. An improvement of roughly one order of magnitude is seen in the error profile of all methods except HF, for which the error was already very small. The enhanced performance of the EXX approximation suggests that, for a given threshold used to solve the OEP equation, a higher plane-wave cutoff helps to improve the accuracy of the EXX potential.
For the OEP methods, the extra source of error related to the evaluation of the local KS potential via the OEP equation seems to have a small impact. This is evident from the good agreement observed between OEP25 and PBE025 in Figs. \ref{fig:Figure_H2O_methods}$(b)$ and \ref{fig:Figure_H2O_methods}$(c)$. Only small irregularities in the error of the force can be seen with OEP25. 

To investigate whether the high accuracy we observe on molecules also extends to solids, we have performed the same analysis of forces on the $\alpha$-quartz phase of SiO$_2$. For this solid that belongs to the P3$_{221}$ space group, cell parameters and atomic positions have first been relaxed at the PBE level using a uniform 333 Monkhorst-Pack grid of \textbf{k} points and a cutoff of 80 Ry. On the optimized structure, one of the silicon ions has then been displaced away from its equilibrium position along the Cartesian x-axis, and the change in the force exerted on this ion has been monitored using PBE, PBE025 and OEP25 at 80 Ry and 200 Ry. The data obtained are presented in Fig. \ref{fig:Figure_SiO2_methods}. 

Similarly to the conclusions drawn from Fig. \ref{fig:Figure_H2O_methods}, we find on silica that PBE025 performs better than PBE at fixed planewave cutoff. The overall error on F$_{\rm A}$ forces is, however, smaller on SiO$_2$ as compared to H$_2$O. This renders the irregularities in the error profile of OEP25 more apparent, despite their amplitudes being very small, about $5 \times 10^{-6}$ Ha/\AA\, at 80 Ry. Given the magnitude of the force exerted on ion Si$^{(1)}$ in Fig. \ref{fig:Figure_SiO2_methods}$(a)$, the errors observed here are not expected to be of practical importance. As already noticed on H$_2$O, using a larger cutoff leads to a decrease of the error in analytical forces for all methods. The OEP25 irregularities are also dampened, reducing from $5 \times 10^{-6}$ Ha/\AA\, at 80 Ry to $2 \times 10^{-6}$ Ha/\AA\, at 200 Ry. 
The different results obtained on SiO$_2$ confirm a good accuracy of the OEP forces for periodic solids. We have been able to identify two main sources of errors in the analytical forces. The first error is method dependent. Different functionals may need different cutoffs to be fully converged. The second error is specific to the OEP method and is related to the accuracy of the local potential generated from the iterative solution of the OEP equation. Both of these errors can be reduced by increasing the plane-wave cutoff and by improving the convergency thresholds of the OEP and KS equations. 
\section{\label{sec:Part5} Applications}
Given the high accuracy seen in the previous section, the calculated OEP forces can be exploited to compute vibrational frequencies. In this section, we will calculate the OEP vibrational frequencies of different molecular and solid state systems, and compare them to the ones predicted by the corresponding methods that use a nonlocal exchange potential. 

The three systems investigated are the water molecule H$_2$O, diamond, and the $\alpha$-quartz phase of SiO$_2$. For H$_2$O, we used the same functionals as tested in Sec. III, i.e. PBE, HF, EXX, PBE025, and OEP25. For diamond and SiO$_2$, only PBE, PBE025 and OEP25 have been used. For each functional we first relaxed the structure. The vibration modes were subsequently determined using the Phonopy Python package \cite{Chaput2023,Togo2023}. Considering the atomic positions and the existing symmetries in the system, this code is able to generate several relevant supercell configurations through slight displacement of the elements. After computation of the analytical forces of each supercell, the existing vibration modes in the system can be predicted by Phonopy and their frequency calculated by finite difference of the forces using the central derivative formula.

\subsection{H$_2$O}
The O-H bond distance \text{d(O-H)}, the bond angle \text{$\Theta$(H-O-H)}, and the vibrational frequencies of the infrared active modes of H$_2$O are presented in Table \ref{tab:Tab_H2O}. The data have been obtained using a simulation box size of 25 Bohr and a plane-wave cutoff of 80 Ry. The accuracy achieved at the end of the optimization procedure is excellent for all approximations. It has been possible to converge the O-H bond distance below $1\times10^{-4}$ \AA , the bond angle below $1\times10^{-2}$ degree, and the vibrational frequencies below 1 cm$^{-1}$. We also tried increasing the cutoff from 80 to 200 Ry without noticing any changes in the results. This confirms that the errors in the forces noted at 80 Ry in Sec. III are sufficiently small to be of no relevance for an accurate determination of equilibrium geometries and vibrational frequencies.
\begin{figure}[t]
\centering
\includegraphics[scale=0.37,angle=0]{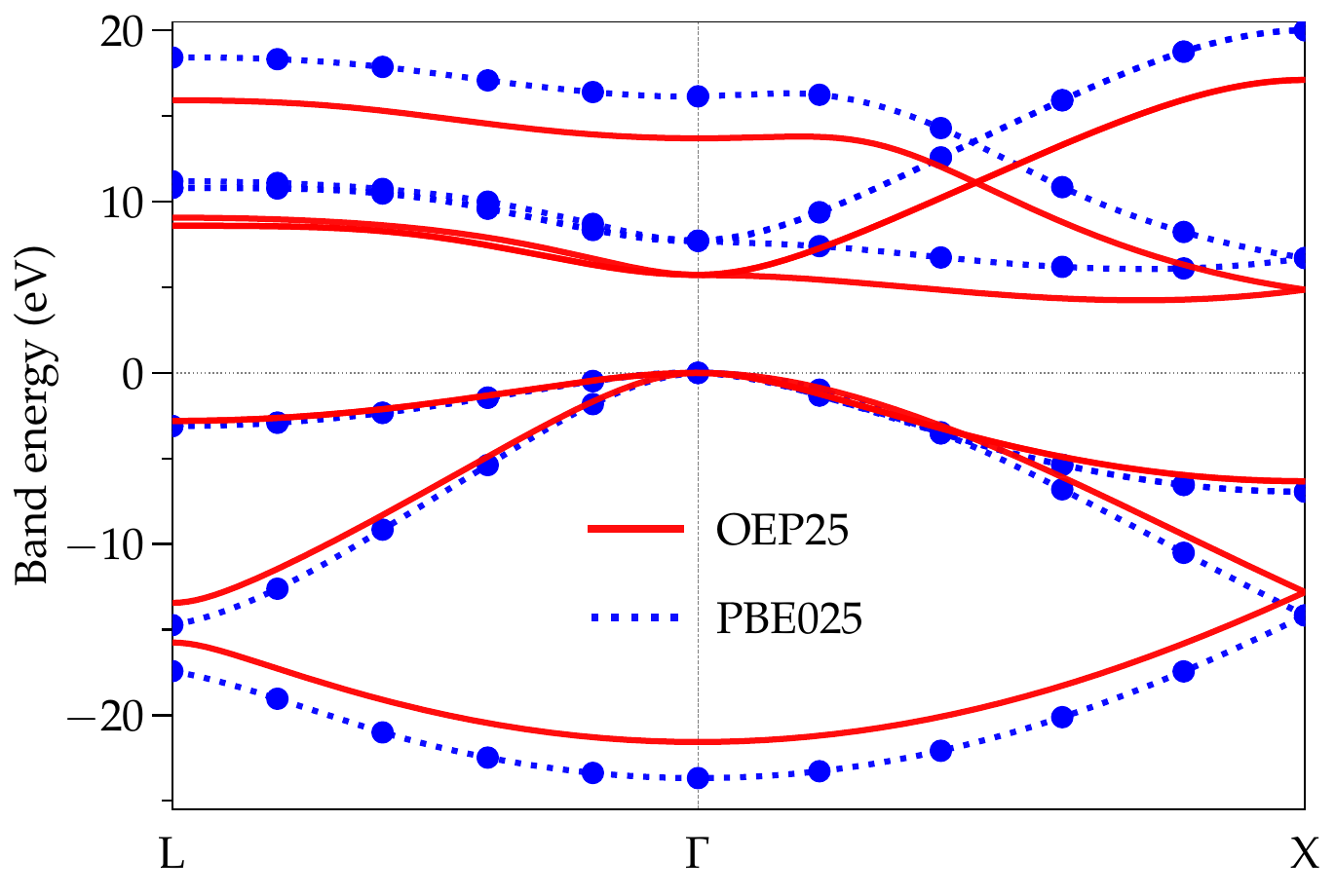}
\caption{Band structure plot of diamond obtained along the high symmetry path L - $\Gamma$ - X. Data have been obtained using PBE025 and OEP25. The PBE025 bands have been spline-interpolated between the dots.}
\label{fig:Figure_bands_C}
\end{figure}

If we compare the structural parameters from each method, we see that the equilibrium geometry predicted by OEP25 is consistent with that of PBE025. The good agreement between these two methods also extends to the vibrational frequencies, with PBE025 and OEP25 returning similar values for the three infrared active modes, namely the in-plane scissoring $\delta_{\rm H-O-H}$, the symmetric stretching $\nu^{s}_{\rm O-H}$, and the asymmetric stretching $\nu^{as}_{\rm O-H}$. On the other hand, small differences can be observed in the structural parameters obtained by the EXX and HF methods. Although very close in geometry, the changes in the O-H bond length and bond angle are sufficient to affect the vibrational frequencies. Compared to HF, EXX returns a lower frequency for the deformation mode, but higher for the two elongation modes. This change is consistent with the evolution of the structural parameters between the two methods as EXX predicts more rigid bonds and a looser angle than HF. The evolution of the structural parameters between EXX and its nonlocal exchange counterpart method HF also agrees with the results presented by Wu et al. \cite{Wu2005}. It confirms that, despite including a maximal fraction of local OEP exact exchange, the EXX method is able to mimic well HF performance. 

Compared to reference CCSD(T)/TZ(2df,2pd) \cite{Thomas1993} and harmonic experimental \cite{Clabo1988} data, PBE025 and OEP25 are the only two methods returning appropriate structural and vibrational properties for the isolated water molecule. 
\begin{figure}[b]
\begin{center}
\includegraphics[scale=0.37,angle=0]{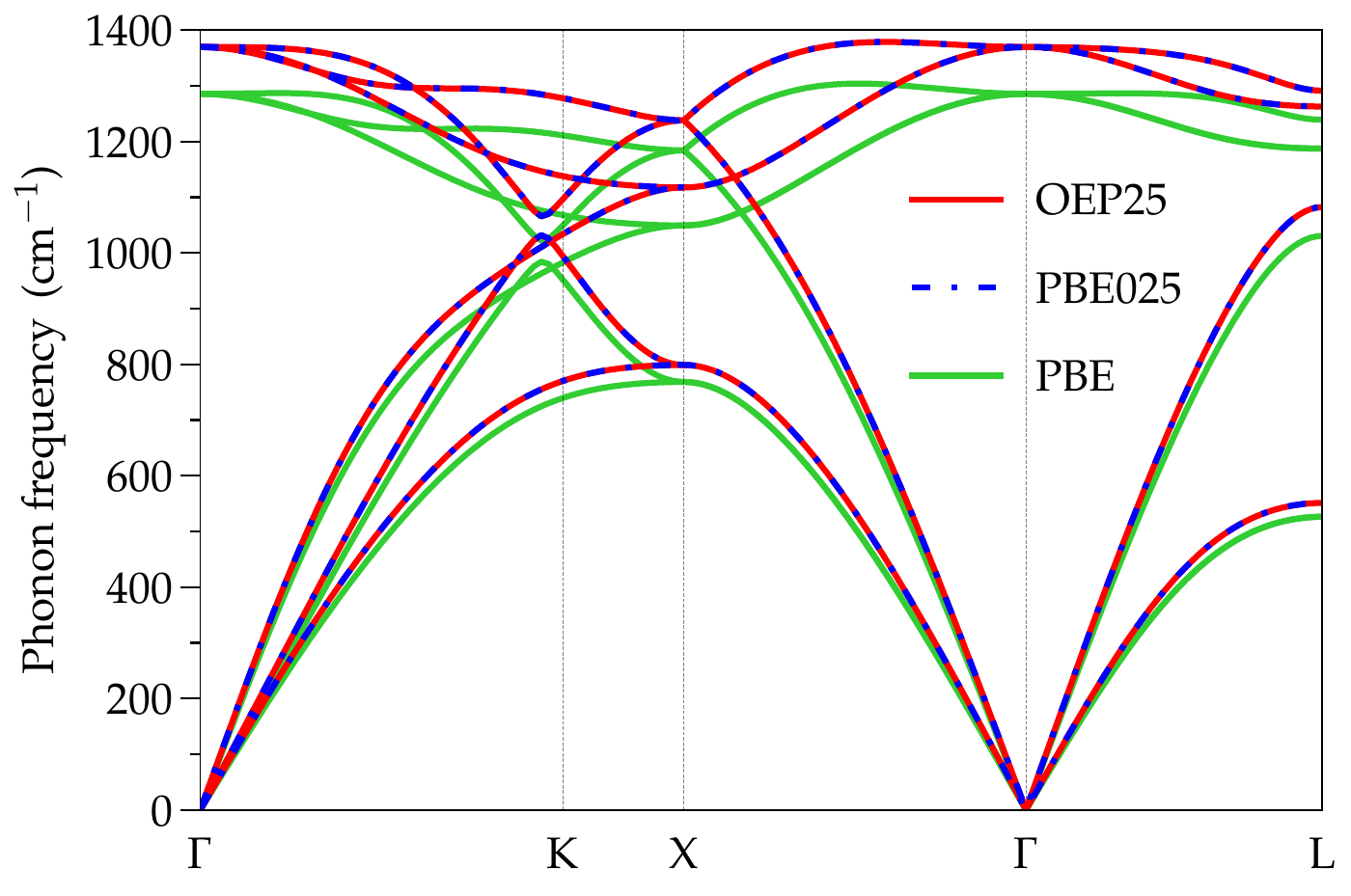}
\caption{\label{fig:Figure_C2_ALL} Phonon dispersion of diamond. The nonlocal PBE025 functional is compared to its local OEP25 counterpart and to PBE.}
\end{center}
\end{figure}
\begin{table*}[ht!]
\caption{\label{tab:Tab_SiO2_structure} Structural parameters of $\alpha$-SiO$_2$ according to different functionals. The internal parameters x, y, and z given describe the position of oxygen and silicon ions within the unit cell. The Si-O bond distances, the Si-O-Si bond angle and the O-Si-O bond angles are also given. For every methods considered, experimental unit cell parameters have been used ($a=4.916$ \AA, $c=5.405$ \AA) \cite{Levien1980}.
}
\vspace{5pt}
\centering
\begin{tabular}{M{2.5cm} M{2.5cm} M{2.5cm} M{2.5cm} M{2.5cm} N}
    \toprule
    \textbf{} & \textbf{PBE} & \textbf{PBE025} & \textbf{OEP25} & \textbf{Expt \cite{Levien1980}} &\\
    \hline
    \textbf{x(Si)} & 0.4685 & 0.4727 & 0.4727 & 0.4697 &\\ 
    \textbf{x(O)} & 0.4120 & 0.4147 & 0.4148 & 0.4135 &\\
    \textbf{y(O)} & 0.2694 & 0.2623 & 0.2623 & 0.2669 &\\
    \textbf{z(O)} & 0.1172 & 0.1235 & 0.1235 & 0.1191 &\\
    \\[-0.2cm]
    \textbf{d(Si-O) (\AA)} & 1.612 & 1.599 & 1.599 & 1.605 &\\
     & 1.617 & 1.603 & 1.603 & 1.614 &\\
    \\[-0.2cm]
    \textbf{$\Theta$(Si-O-Si) (°)} & 142.7 & 145.3 & 145.3 & 143.7 &\\
    \\[-0.2cm]
    \textbf{$\Theta$(O-Si-O) (°)} & 108.7 & 108.9 & 108.9 & 108.8 &\\

     & 108.7 & 108.9 & 108.9 & 109.0 &\\

     & 109.1 & 109.2 & 109.2 & 109.2 &\\

     & 110.8 & 110.4 & 110.4 & 110.5 &\\
    \toprule
\end{tabular}
\end{table*}
\subsection{Diamond}
Diamond crystallizes in the Fd-3m space group. Thanks to the high symmetry, the atomic positions are fixed within the unit cell. Only the lattice parameter $a$ is left to vary in order to identify the optimized unit cell. We have identified the suitable cell dimensions by monitoring the change of the total energy upon variations of the lattice parameter $a$. The values of 6.700 Bohr for both OEP25 and PBE025, and 6.742 Bohr for PBE are optimal. 

We first calculated the electronic band structure along the L - $\Gamma$ - X high symmetry path within PBE025 and OEP25 (see Fig.~\ref{fig:Figure_bands_C}). The OEP25 bands are easy to generate using post-processing tools since the local exchange potential is ${\mathbf k}$-independent. For PBE025 we only have the band energies on the ${\mathbf k}$-point grid used in the self-consistent calculation. The bands have, therefore, been spline-interpolated between these points. The band structures are clearly different in the two approximations. The first conduction band is shifted by approximately 2 eV in PBE025. This difference is expected and related to the derivative discontinuity within the OEP method \cite{Li1991,Perdew1985}. Adding the derivative discontinuity correction to the OEP25 result returns a gap in very good agreement with PBE025, with an energy difference of 0.02 eV.

We calculated the phonon dispersion along a high symmetry path in the Brillouin Zone with Phonopy using a single 2$\times$2$\times$2 supercell containing 64 carbon atoms. 
A uniform 222 Monkhorst-Pack grid of \textbf{k} points and a plane wave cutoff of 100 Ry have been employed. The results are presented in Fig. \ref{fig:Figure_C2_ALL}. Despite the use of the frozen phonon approach, our PBE results agree well with the data presented by Mounet et al. \cite{Mounet2005} using DFPT. We also notice a very good agreement between the PBE025 and OEP25 frequencies. As expected, both methods return higher frequencies than PBE because the effect of exact exchange is well known to strengthen bonds. This case study proves that accurate phonons can be calculated using the OEP method.
\begin{table*}[ht!]
\caption{\label{tab:Tab_SiO2_phonons} Phonon frequencies calculated at the $\Gamma$-point of $\alpha$-SiO$_2$ on the relaxed structures presented in Table \ref{tab:Tab_SiO2_structure} for different functionals. Frequencies are given in cm$^{-1}$. While the acoustic sum rule is applied, the LO-TO splitting is only considered in experimental results \cite{He2014}. Phonon frequencies affected by LO-TO splitting are underlined in the table.
}
\vspace{5pt}
\centering
\begin{tabular}{M{3.5cm} M{2.5cm} M{2.5cm} M{2.5cm} M{2.5cm} N}
    \toprule
    \textbf{} & \textbf{PBE} & \textbf{PBE025} & \textbf{OEP25} & \textbf{Expt \cite{He2014}} &\\
    \hline
    \textbf{E$_{\rm \textbf{u}}$(TO1)} & 140 & 67 & 67 & 133 &\\
    \textbf{E$_{\rm \textbf{u}}$(LO1)} & \underline{140} & \underline{67} & \underline{67} & 133 &\\
    \textbf{A$_{\rm \textbf{1}}$(1)} & 211 & 167 & 167 & 219 &\\
    \textbf{E$_{\rm \textbf{u}}$(TO2)} & 263 & 242 & 242 & 269 &\\
    \textbf{E$_{\rm \textbf{u}}$(LO2)} & \underline{263} & \underline{242} & \underline{242} & 269 &\\
    \textbf{A$_{\rm \textbf{1}}$(2)} & 351 & 354 & 354 & 358 &\\
    \textbf{A$_{\rm \textbf{2}}$(TO1)} & 352 & 363 & 364 & 361 &\\
    \textbf{E$_{\rm \textbf{u}}$(TO3)} & 386 & 388 & 388 & 394 &\\
    \textbf{E$_{\rm \textbf{u}}$(LO3)} & \underline{386} & \underline{388} & \underline{388} & 402 &\\
    \textbf{E$_{\rm \textbf{u}}$(TO4)} & 442 & 446 & 447 & 453 &\\
    \textbf{E$_{\rm \textbf{u}}$(LO4)} & \underline{442} & \underline{446} & \underline{447} & 512 &\\
    \textbf{A$_{\rm \textbf{1}}$(3)} & 452 & 453 & 453 & 469 &\\
    \textbf{A$_{\rm \textbf{2}}$(TO2)} & 484 & 489 & 489 & 499 &\\
    \textbf{E$_{\rm \textbf{u}}$(TO5)} & 677 & 696 & 696 & 698 &\\
    \textbf{E$_{\rm \textbf{u}}$(LO5)} & \underline{677} & \underline{696} & \underline{696} & 701 &\\
    \textbf{A$_{\rm \textbf{2}}$(TO3)} & 756 & 782 & 783 & 778 &\\
    \textbf{E$_{\rm \textbf{u}}$(TO6)} & 776 & 797 & 797 & 799 &\\
    \textbf{E$_{\rm \textbf{u}}$(LO6)} & \underline{776} & \underline{797} & \underline{797} & 812 &\\
    \textbf{E$_{\rm \textbf{u}}$(TO7)} & 1042 & 1085 & 1085 & 1066 &\\
    \textbf{E$_{\rm \textbf{u}}$(LO7)} & \underline{1042} & \underline{1085} & \underline{1085} & 1227 &\\
    \textbf{A$_{\rm \textbf{2}}$(TO4)} & 1050 & 1092 & 1093 & 1072 &\\
    \textbf{A$_{\rm \textbf{1}}$(4)} & 1061 & 1101 & 1102 & 1082 &\\
    \textbf{E$_{\rm \textbf{u}}$(TO8)} & 1133 & 1181 & 1182 & 1158 &\\
    \textbf{E$_{\rm \textbf{u}}$(LO8)} & \underline{1133} & \underline{1181} & \underline{1182} & 1155 &\\
    \toprule
\end{tabular}
\vspace{1pt}
\end{table*}
\subsection{SiO$_2$}
The last system we considered is the $\alpha$-quartz phase of SiO$_2$. Unlike diamond, $\alpha$-quartz silica has polar bonds. Therefore, correct phonons can only be obtained if the LO-TO splitting is explicitly calculated. To evaluate the LO-TO splitting, knowledge of the dielectric tensor and Born effective charges of all symmetry-inequivalent ions within the unit cell is required \cite{Baroni2001,Gonze1997}. However, since our present objective is to compare the performance of the OEP25 functional with respect to PBE025, we have neglected the LO-TO splitting correction when computing the phonon modes of SiO$_2$. 

For this study, we have used experimental unit cell parameters \cite{Levien1980} and then relaxed the atomic positions for each functional. We have used a uniform 333 Monkhorst-Pack dense grid of \textbf{k} points and a plane-wave cutoff of 100 Ry. The structural parameters are presented in Table \ref{tab:Tab_SiO2_structure}. The internal coordinates x(Si), x(O), y(O), and z(O) describe, along with the symmetries of the crystal, the position of each ion within the cell. Given the complexity of the SiO$_2$ unit cell, 
it is remarkable that PBE025 and OEP25 produce similar structural parameters. These two methods give tighter bonds and looser angles than PBE. On the other hand, experimental structure parameters appear to be located halfway between these methods. 

We then calculated the phonon modes at the $\Gamma$-point by generating a total of nine different 1$\times$1$\times$1 supercells with Phonopy. 
The list of phonon modes and their associated frequencies are presented in Table \ref{tab:Tab_SiO2_phonons}. For each mode, we find an excellent agreement between PBE025 and OEP25. As expected, by including a fraction of exact exchange, the frequencies obtained are in general higher in energy than that of PBE. For the TO modes located over 1000 cm$^{-1}$, experimental frequencies are located halfway between PBE and PBE025/OEP25 results, which might indicate that the true fraction of exact exchange to include for PBE0 should actually be lower than the standard 25\%. The comparison with experimental data also show for which LO modes a significant shift of the frequencies we calculated can be expected. 

This comprehensive analysis on SiO$_2$ $\alpha$-quartz confirms that our implementation of OEP forces can be used for accurate studies of complex systems.

\section{\label{sec:Part6} Conclusions and outlook}
Achieving full self-consistency with orbital-dependent xc functionals requires the solution of the OEP equation. Although numerically challenging, the OEP represents a simplification over the use of nonlocal and energy-dependent potentials. Being formulated within the KS-DFT framework, it is, for example, easy to combine the OEP with existing codes for excited state properties (e.g. $GW$) or lattice dynamics.

In this work, we have shown that OEP forces, within hybrid functionals, can be computed with a numerical accuracy similar to that obtained with commonly used functionals in DFT. However, we also showed that special care is needed when employing nonlocal pseudopotentials. Since the OEP is based on a constrained optimization, an extra force term needs to be added to the standard Hellmann-Feynman expression for the forces. We implemented this term within the \texttt{ACFDT} package of the \texttt{QUANTUM ESPRESSO} distribution, that already computes the OEP potential. 
This allowed us to calculate forces, relax geometries, and determine phonon frequencies for a number of molecules and solids. Our different studies show that the local OEP exchange potential is a good approximation to the nonlocal exchange potential, being able to produce almost identical equilibrium structures and phonon frequencies. 

The high numerical accuracy we have obtained with the OEP applied to hybrid functionals paves the way for determining the OEP forces also with more advanced functionals, such as those based on MBPT. Furthermore, our work provides a first step towards the calculation of phonon spectra and electron-phonon couplings within DFPT, using an advanced treatment of exchange and correlation.
\begin{acknowledgements}
The authors would like to thank Michele Casula and Lorenzo Paulatto for helpful discussions. The work was partially performed using HPC resources from GENCI-TGCC/CINES/IDRIS (Grant No. A0150914650). 
\end{acknowledgements}

\providecommand{\noopsort}[1]{}\providecommand{\singleletter}[1]{#1}%

\end{document}